\documentclass[%
 aip,
 jmp,%
 amsmath,amssymb,
 reprint,%
]{revtex4-1}

\usepackage{graphicx}
\usepackage{dcolumn}
\usepackage{bm}
\usepackage{multirow}
\usepackage{color} 

\begin{document}

\preprint{AIP/123-QED}

\title[Philosometrics]{Philosometrics}

\author{Renato Fabbri}%
 \homepage{www.estudiolivre.org/el-user.php?view\_user=gk}
 \email{renato.fabbri@gmail.com}
  \affiliation{ 
Instituto de F\'isica de S\~ao Carlos, Universidade de S\~ao Paulo (IFSC/USP)
}%

\author{Osvaldo N. Oliveira Jr.}
  \homepage{www.polimeros.ifsc.usp.br/professors/professor.php?id=4}
  \email{chu@ifsc.usp.br}
 \altaffiliation[Also at ]{IFSC-USP}

\author{Luciano da Fontoura Costa}
  \homepage{http://cyvision.ifsc.usp.br/~luciano/}
  \email{ldfcosta@gmail.com}
  \altaffiliation[Also at ]{IFSC-USP}

\date{\today}

\begin{abstract}
The hypothesis that philosophy is driven by difference/innovation is
checked in a quantitative manner. This was performed by assigning
grades to eight main philosophical features with respect to seven
prominent philosophers, which allowed sound concepts from multivariate
statistics and pattern recognition to be applied. A number of insights
could then be inferred about the way in which philosophy has been
developed. For instance, the evolution of philosophy can be
represented as a trajectory on a plane (two instead of the eight
original dimensions) whose axes are defined by a linear combination of
the philosophical characteristics considered.  In addition, all the
philosophical moves have been verified to oppose the prevailing
philosophical state.  We also identified an intense and progressive
trend toward dialectics.
\end{abstract}

\maketitle

\begin{quotation}
`One cannot conceive anything so strange and so implausible that it
has not already been said by one philosopher or another.'

\emph{- R. Descartes}
\end{quotation}

\section{\label{sec:level1}Introduction}

We wished to start this work in a different way. So, we chose to begin
it by stating this objective.  By doing so, we immediately incurred
into a paradox, for most authors want to be different from the outset
and that objective would ultimately imply in being no different at
all.  In other ways, many authors have tried to distinguish themselves
by being different.  We believe philosophers are no different in this
respect, i.e.\ they inherently want not to be equal and therefore tend
to develop new philosophical views and paradigms.  This was the
original motivation for the present work.

The emphasis on the quest for difference in philosophy has been
expressed by Gilles Deleuze~\cite{Deleuze}, but can also be
immediately related to other previous approaches such as Ferdinand de
Saussurre's principle~\cite{Saussure}, in which concepts (words) tend
to be different in the sense of meaning distinct things.  The paradigm
of difference is particularly promising because it is immediately
related to the own dynamics of philosophical evolution along time.  In
other words, what are the forces that drive philosophical changes and
innovation?  Could dialectics account for some of the main
philosophical moves?

In order to try and answer these questions in a more objective way, we
resorted to a quantitative approach which turned out to provide a
number of remarkable spin-offs and insights.  More specifically, we
started by identifying prominent philosophers along the history of
western philosophy, a set of main philosophical issues, and then
assign grades to each of these issues for the chosen philosophers.  It
should be stressed that we chose a reduced set of philosophers for the
sake of simplicity and clarity.  Though limited by the arbitrary
procedure with which the grades were assigned, the quantification of
the main philosophers (and consequently of their views)
characteristics paved the way to the application of sound concepts and
methods from multivariate statistics~\cite{Papoulis, Wichern,
Therrien} and pattern recognition~\cite{Duda, Costa}.  We checked for
correlations between the characteristics, and then applied principal
component analysis (PCA)~\cite{Costa} to investigate the dispersion of
the philosophical principles.  We also considered the time sequence of
the philosophers, which allowed us to define respective trajectories
of philosophical evolution.  By proposing indices to quantify the
degree of opposition, skewness and dialectics along the dynamics of
philosophical evolution, we observed surprising results.

This work starts by presenting the methodology adopted and then
presents the definition of the eight philosophical characteristics
chosen, followed by the description of the results and discussion.

\section{Mathematical Description}

The choice of philosophers is inherently important for the
time-evolution analysis, as each philosophical move and its
characteristics (i.e.\ opposition and skewness) are defined by each
pair of subsequent philosophers along time.  We decided to choose the
philosophers taking into account their historical importance and
visibility.  A sequence $S$ of $P$ philosophers along a given period
of time would incorporate the $P$ most prominent and visible
philosophers in that interval.  The use of such a criterion to build
the time-sequence for the philosophers implies in not necessarily
uniform time-intervals between each pair of subsequent entries.

The set of $C$ measurements used to characterize the philosophers
define a $C-$dimensional feature space which will be henceforth
referred to as the \emph{philosophical space}.  The characteristic
vector $\vec{v_i}$ of each philosopher $i$ defines a respective
\emph{philosophical state} in the philosophical space.  Given a set of
$P$ philosophers, the \emph{average state at time $i$}, $i \leq P$, is
defined as

\begin{equation}
  \vec{a_i} = \frac{1}{i}\sum_{k=1}^i\vec{v_k}.
\end{equation}

The \emph{opposite state} of a given philosophical state $\vec{v_i}$
is defined as:

\begin{equation}
  \vec{r_i} = \vec{v_i} + 2(\vec{a_i} - \vec{v_i})  = 2 \vec{a_i} - \vec{v_i}
\end{equation}

The \emph{opposition vector} of philosophical state $\vec{v_i}$ is
given us by $\vec{D_i}=\vec{r_i} - \vec{v_i}$.  The \emph{opposition
amplitude} of that same state is defined as $|| \vec{D_i} ||$.

An \emph{emphasis move} taking place from the philosophical state
$\vec{v_i}$ is any displacement from $\vec{v_i}$ along the direction
$-\vec{r_i}$.  A \emph{contrary move} from the philosophical state
$\vec{v_i}$ is any displacement from $\vec{v_i}$ along the direction
$\vec{r_i}$.

Given a time-sequence $S$ of $P$ philosophers, the \emph{philosophical
move} implied by two successive philosophers $i$ and $j$ corresponds
to the $\vec{M_{i,j}}$ vector extending from $\vec{v_i}$ to
$\vec{v_j}$, i.e.

\begin{equation}
  \vec{M_{i,j}} = \vec{v_j} - \vec{v_i}
\end{equation}

In principle, an innovative or differentiated philosophical move would
be such that it departs substantially from the current philosophical
state $\vec{v_i}$.  We decided to decompose innovation moves into two
main subtypes: opposition and skewness.

The \emph{opposition index} $W_{i,j}$ of a given philosophical move
$\vec{M_{(i,j)}}$ is defined as

\begin{equation}
  W_{i,j} = \frac{\left< \vec{M_{i,j}}, \vec{D_i}\right>}{||\vec{D_i}||^2}. 
\end{equation}

This index quantifies the intensity of opposition of that respective
philosophical move, in the sense of having a large projection along
the vector $\vec{D_i}$.  It should also be noticed that the repetition
of opposition moves lead to little innovation, as it would imply in an
oscillation around the average state.

The \emph{skewness index} $s_{i,j}$ of that same philosophical move is
the distance beetwen $\vec{v_j}$ and the line $L_i$ defined by the
vector $\vec{D_i}$, and therefore quantifies how much the new
philosophical state departs from the respective opposition move.
Actually, a sequence of moves with zero skewness would represent more
trivial oscillations within the opposition line $L_i$.

We also suggest an index to quantify the dialectics between a triple
of successive philosophers $i, j$ and $k$.  More specifically, the
philosophical state $\vec{v_i}$ is understood as the \emph{thesis},
the state $j$ is taken as the \emph{antithesis}, with the synthesis
being associated to the state $\vec{v_k}$~\cite{Papineau}.  The
hypothesis that $k$ is the consequence, among other forces, of a
dialectics between the views $\vec{v_i}$ and $\vec{v_j}$ can be
expressed by the fact that the philosophical state $\vec{v_k}$ be
located near the middle line $ML_{i,j}$ defined by the thesis and
antithesis (i.e.\ the points which are at an equal distance to both
$\vec{v_i}$ and $\vec{v_j}$) relatively to the opposition amplitude
$|| \vec{D_i} ||$.

Therefore, the \emph{counter-dialectic index} is defined as 

\begin{equation}
\rho_{i \rightarrow k} = d_{i \rightarrow k}/ || \vec{M_{i,j}}||, 
\end{equation}

where $d_{i \rightarrow k}$ is the distance between the philosophical
state $\vec{v_k}$ and the middle-line $ML_{i,j}$ between $\vec{v_i}$
and $\vec{v_j}$~\footnote{In higher dimensional philosophical spaces,
the middle-hyperplane defined by the points which are at equal
distances to both $\vec{v_i}$ and $\vec{v_j}$ should be used instead
of the middle-line.}.  Note that $0 \leq d_{i \rightarrow k} \leq
1$.  The choice of counter-dialectics instead of dialectics is justified
to maintain compatibility with the use of a distance from point
to line as adopted for the definition of skewness.
\begin{figure}
        \begin{center}
                \includegraphics[width=0.35\textwidth]{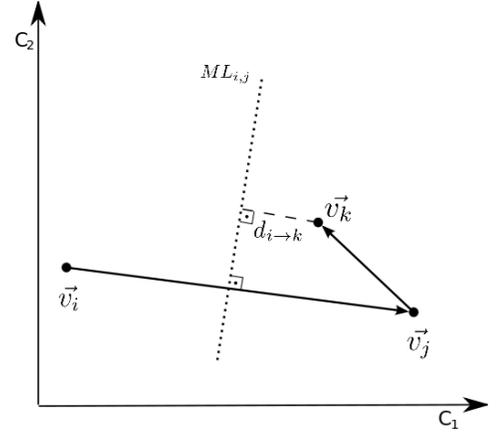}
        \end{center}
        \caption{\it Graphical representation of the quantification of dialectics.}
        \label{fig.1}
\end{figure}

\begin{figure}
        \begin{center}
                \includegraphics[width=0.35\textwidth]{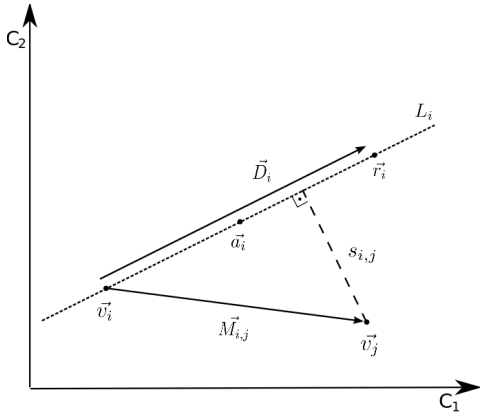}
        \end{center}
        \caption{\it Graphical representation of the measures derived from a \emph{philosophical move}.}
        \label{fig.1}
\end{figure}

\section{Philosophical Axes}
We derived eight variables corresponding to some of the most recurrent
philosophical issues~\cite{Russel,Papineau,Deleuze2}.  Each of
these variables, which define a respective axis in the philosophical
space, are briefly described in the following.

{\bf \em{ Rationalism - Empirism}} (\emph{R-E}): the rationalists
claim that the human acquaintance of knowledge/concepts is
significantly independent of sense experience. Empiricists understand
sense experience as the main way to gain knowledge/concepts.
Frequently, rationalists take the view that the world is affected by
intrinsic properties of the human brain, in contrast to the empiricist
approach where the world would imprint itself onto our minds.

{\bf \em{ Essence - Existence}} (\emph{E-E}): An existence-based
understanding of the world has its basis on the fact that things are
as an existent unit. Essence focuses on a substance
(e.g. intellectual) that precedes existence itself.

{\bf \em{ Monism - Dualism}} (\emph{M-D}): Dualism requires the
division of the human person into two or more domains, such as matter
and soul. Monism is based on a unique "category of being".

{\bf \em{ Theocentrism - Anthropocentrism}} (\emph{T-A}): In
theocentrism, God is the most important thing in the universe.  The
anthropocentric view has man as prevalent.

{\bf \em{ Holism - Reductionism}} (\emph{H-R}): Reductionism attempts
to explain the world in terms of simple components and its emerging
properties.  Holists focus on the fact that the whole is more than its
constitutive parts.

{\bf \em{ Deductionism - Phenomenology}} (\emph{D-P}): Phenomenology
relies on systematic reflection of consciousness and what happens in
conscious acts.  Deductionism is based on deriving conclusions from
axiomatic systems.  {\bf \em{ Determinism - Free Will}} (\emph{D-F}):
Free will assumes that humans make choices and these are not
predetermined.  Determinism understands that every event is fatidic,
e.g., perfectly determined by prior states.

{\bf \em{ Naturalism - Mechanism}} (\emph{N-M}): Methodological
naturalism is the thinking basis of modern science, i.e. hypotheses
must be argued and tested in terms of natural laws. Mechanism attempts
to build explanation using logic-mathematical processes.

\section{Results and Discussion}

A set of seven philosophers were chosen spanning the period from
Classical Greece until contemporary times, and ordered chronologically
as: Plato, Aristotle, Descartes, Espinoza, Kant, Nietzsche, and
Deleuze.  The quantification of the eight philosophical
characteristics was performed jointly by two of the authors of this
article and is shown in Table \ref{tab:tableA}.

\begin{table}
\caption{\label{tab:tableA}Quantification of the
eight philosophical characteristics for each of the seven philosophers.  }

\begin{ruledtabular}
\begin{tabular}{|l||c|c|c|c|c|c|c|c|}

Philosophers & R-E & E-E & M-D & T-A & H-R & D-P & D-F & N-M \\ \hline

Plato  & 3.0 &   3.5 &   9.0  &   5.  &   4.5 &   3.5 &   5.0 &   4.5 \\

Aristotle & 8.0 &   7.5 &   7.0  &   5.5 &   7.5 &   8.0 &   2.5 &   2.5 \\

Descartes & 1.5 &   2.5 &   9.0  &   6.5 &   7.0 &   2.5 &   7.5 &   7.5 \\

Espinoza     & 8.0 &   2.0 &   1.0  &   5.0 &   2.0 &   3.0 &   1.0 &   1.0 \\

Kant      & 7.0 &   2.5 &   8.5  &   6.5 &   4.5 &   3.5 &   7.5 &   5.0 \\

Nietzsche & 7.5 &   9.0 &   1.0  &   9.0 &   5.0 &   8.0 &   1.0 &   1.5 \\

Deleuze   & 5.5 &   7.5 &   1.0  &   8.  &   2.5 &   5.5 &   5.0 &   6.0 \\
\end{tabular}
\end{ruledtabular}
\end{table}

This dataset defines an 8-dimensional philosophical space.  The
Pearson correlation coefficients between the eight philosophical
characteristics chosen are presented in Table \ref{tab:tableB}.  The
coefficients with absolute value larger than 0.35 are emphasized.
Strong positive or negative correlations can be observed for several
pairs of characteristics.  For instance, the fact that a Pearson
correlation coefficient of $-0.80$ was obtained for the pair of
characteristics \emph{R-E} and \emph{N-M} indicates that philosophers
who are rationalists strongly tend to be also mechanicists.  An even
stronger correlation of $0.94$, now positive, is observed between
\emph{E-E} and \emph{D-P}, suggesting that existentialists also tend
to be phenomenologists, as could be expected.  Other strong
correlations were observed, including a Pearson coefficient of $0.92$
between \emph{D-F} and \emph{N-M}.  Also interesting is the relatively
high correlation between \emph{M-D} and \emph{D-F}, which seems to be
directly implied by religious background.

\begin{table}\footnotesize
\caption{\label{tab:tableB}Pearson correlation coefficients between the eight philosophical characteristics.  The entries with absolute
value larger or equal than 0.35 have been emphasized.}

\begin{ruledtabular}
\begin{tabular}{|c||c|c|c|c|c|c|c|c|}

- & R-E & E-E & M-D & T-A & H-R & D-P & D-F & N-M \\ \hline
R-E & 1.00 & {\bf \emph{  0.37}} & {\bf \emph{  -0.55}} & 0.11 & -0.26 & {\bf \emph{  0.56}} & {\bf \emph{  -0.67}} & {\bf \emph{  -0.80}} \\
E-E & -    & 1.00 & {\bf \emph{  -0.48}} & {\bf \emph{  0.67}} & 0.16 & {\bf \emph{  0.94}} & {\bf \emph{  -0.47}} & -0.27 \\
M-D & -    & -    & 1.00 & {\bf \emph{  -0.49}} & {\bf \emph{  0.64}} & {\bf \emph{  -0.37}} & {\bf \emph{  0.70}} & {\bf \emph{  0.53}} \\
T-A & -    & -    & -    & 1.00 & -0.05 & {\bf \emph{  0.49}}  & -0.06 & 0.09 \\
H-R & -    & -    & -    & -    & 1.00 & 0.29 & 0.23 & 0.22 \\ 
D-P & -    & -    & -    & -    & -    & 1.00 & {\bf \emph{  -0.57}}  & {\bf \emph{  -0.47}} \\
D-F & -    & -    & -    & -    & -    & -    & 1.00 & {\bf \emph{  0.92}} \\
N-M & -    & -    & -    & -    & -    & -    & -    & 1.00 \\

\end{tabular}
\end{ruledtabular}
\end{table}

PCA was applied to this set of data, yielding the new variances given
in Table \ref{tab:tableC} in terms of the percentages of the total
variance.  It is clear that the two first PCA axes account for as much
as $75\%$ of the total variance of the dataset in the original
philosophical space.  It is therefore reasonable to say that the
properties of the philosophers can be well described and analysed
while considering just two dimensions, yielding a planar philosophical
space.

\begin{table}
\caption{\label{tab:tableC}New variances after PCA, in percentages.}

\begin{tabular}{|c||c|}
\hline
Eigenvalue & Value \\ \hline
$\lambda_1$ &  51.26 \% \\
$\lambda_2$ &  23.57 \% \\
$\lambda_3$ &  16.70 \% \\
$\lambda_4$ &  4.76 \% \\
$\lambda_5$ &  2.64 \% \\
$\lambda_6$ &  1.08 \% \\
$\lambda_7$ &    0. \%    \\
$\lambda_8$ &    0. \%    \\
\hline

\end{tabular}
\end{table}

In order to investigate the effect from the unavoidable errors in the
quantification of the philosophical characteristics, we performed 1000
perturbations of the original scores by adding the values -2, -1, 0,
1, and 2 with uniform probability.  Table \ref{tab:tableD} shows the
average and standard deviation of the deviations obtained for each
philosopher considering the original and perturbed positions, as well
as the average and standard deviation of the first 4 eigenvalues
(i.e. the variances of the perturbed configurations).  Interestingly,
these results show relatively small effects of the perturbations on
the PCA projections and variances.  In other words, the assignment of
the original measurements does not seem to be too critical for the
results.

\begin{table}
\caption{\label{tab:tableD}Average and standard deviation of the 
deviations for each philosopher and for the first 
4 eigenvalues.  }

\begin{tabular}{|c||c|c|}
\hline

Philosophers & $\mu_{\Delta}$ & $\sigma_{\Delta}$ \\
\hline
Plato     & 0.9688 & 0.5707 \\
Aristotle & 1.7975 & 1.0990 \\
Descartes & 1.0106 & 0.6179 \\
Espinoza & 2.3583 & 1.9020 \\
Kant & 0.8970 & 0.5071 \\
Nietzsche & 1.2220 & 0.8562 \\
Deleuze & 1.3474 & 0.7541 \\
\hline \hline
Eigenvalues & $\mu_{\Delta}$ & $\sigma_{\Delta}$ \\
\hline
$\lambda_1$ &  0.0449 & 0.0533 \\
$\lambda_2$ &  0.0163 & 0.0372 \\
$\lambda_3$ &  0.0129 & 0.0301 \\
$\lambda_4$ &  0.0323 & 0.0231 \\
\hline

\end{tabular}
\end{table}

It is particularly striking that the seeming complexity of the
philosophical traits quantified in terms of the eight characteristics
boiled down to no more than two dimensions for their representation.
This is a direct consequence of the high Pearson correlation
coefficients between several of those characteristics, meaning that
most of them tend to be related.

Also shown in Table \ref{tab:Deviates} are the normalized weights (as
percentages) of the contributions of each original property on the two
new main axes.  Most of the characteristics contribute almost equally
in defining the two main axes.  Consequently, we now have two
possibilities for representing the philosophical space: to consider
almost equally all these characteristics or to define two new
philosophical features as given by the PCA.

\begin{table}
\caption{\label{tab:Deviates}Percentages of
the contributions from each philosophical characteristic on the two new main axes.  }
\begin{tabular}{|c||c|c|}
\hline
Philosophical & \multirow{2}{*}{$C_1$} & \multirow{2}{*}{$C_2$} \\
Characteristics & & \\
\hline
R-E & 14.60 & 7.30  \\
E-E & 13.44    & 17.25 \\
M-D & 14.67    & 6.31 \\
T-A & 7.87    & 16.46 \\
H-R & 4.57    & 18.48 \\
D-P & 14.37    & 14.87 \\
D-F & 15.96    & 7.76 \\
N-M & 14.48    & 11.59 \\
\hline

\end{tabular}
\end{table}

\begin{figure}
        \begin{center}
                \includegraphics[width=0.45\textwidth]{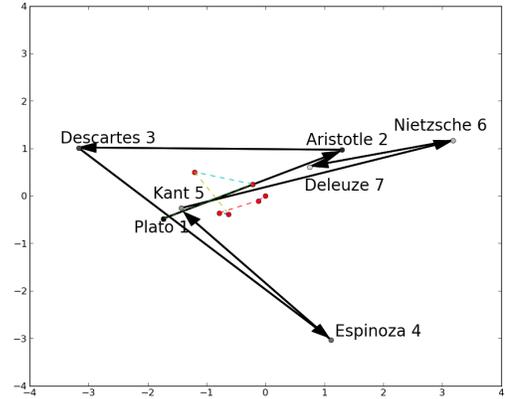}
        \end{center}
        \caption{\it 2-dimensional projected philosophical space.}
        \label{fig:pca}
\end{figure}

\begin{figure}
        \begin{center}
                \includegraphics[width=0.45\textwidth]{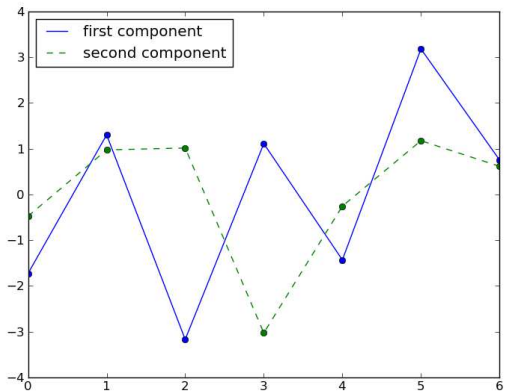}
        \end{center}
        \caption{\it Time evolution of the two principal components.}
        \label{fig:time}
\end{figure}

The 2-dimensional projected space is presented in Figure
\ref{fig:pca}, where the arrows follow the time sequence along with
the chosen philosophers.  Recall that each of these arrows corresponds
to a philosophical move.  Interestingly, the time-averages, also shown
in this figure, exhibit relatively small displacements.
Figure~\ref{fig:time} shows the two principal components in terms of
time-slot.  The first component seems to have a period which is twice
as small as that of the second component.  It is also clear from this
figure that the philosophical states seem to have a period
characterized by nearly the same values at the 1 and 6 time-slots.

Table \ref{tab:tableOI} gives the opposition and skewness indices
for each of the six philosophical moves.

\begin{table}
\caption{\label{tab:tableOI}Opposition and skewness indices for each
of the six philosophical moves.  }

\begin{tabular}{|c||c|c|}
\hline
Philosophical Move & $W_{i,j}$ & $s_{i,j}$ \\
\hline \hline
Plato $\rightarrow$ Aristotle &  1.0 & 0 \\
Aristotle $\rightarrow$ Descartes & 0.8622 & 0.8656 \\
Descartes $\rightarrow$ Espinoza & 0.9803 & 1.4930 \\
Espinoza $\rightarrow$ Kant & 0.5693 & 0.4715 \\
Kant $\rightarrow$ Nietzsche & 0.8021 & 0.8726 \\
Nietzsche $\rightarrow$ Deleuze & 0.3647 & 0.3148 \\
\hline
\end{tabular}
\end{table}

\begin{table}
\caption{\label{tab:tableE} Counter-dialectics index for each
of the five subsequent pairs of philosophical moves.  }

\begin{tabular}{|c||c|}
\hline
Philosophical Triple & $d_{i \rightarrow k}$ \\
\hline \hline
Plato $\rightarrow$ Aristotle $\rightarrow$ Descartes &  0.700 \\
Aristotle $\rightarrow$ Descartes $\rightarrow$ Espinoza & 0.466 \\
Descartes $\rightarrow$ Espinoza $\rightarrow$ Kant & 0.137 \\
Espinoza $\rightarrow$ Kant $\rightarrow$ Nietzsche & 0.048 \\
Kant $\rightarrow$ Nietzsche $\rightarrow$ Deleuze  & 0.015 \\
\hline
\end{tabular}
\end{table}

All the philosophical moves tend to take place according to a
well-defined and intense opposition from the average state.  Also
surprisingly, rather small skewness has been found to underlie most
philosophical moves, meaning that most philosophical moves are driven
almost exclusively by opposition to the current philosophical state.
Remarkable results have been obtained also for dialectics.  We
identified progressively stronger dialectics trends among subsequent
pairs of philosophical moves.

In order to complement our analysis of the relationship between the
philosophers, we applied Wards hierarchical clustering
algorithm~\cite{Duda,Costa} considering the eight original features.
This methodology clusters the individuals progressively taking into
account their intra-cluster dispersion, so that the obtained
dendrogram reflects the similarity between the philosophers.  Two main
groups can be identified, one including Plato, Descartes and Kant, and
the other containing the remainder philosophers.  The former group,
which can also be identified in the left-hand side of the PCA diagram
in Figure~\ref{fig:pca}.

\begin{figure}
        \begin{center}
                \includegraphics[width=0.5\textwidth]{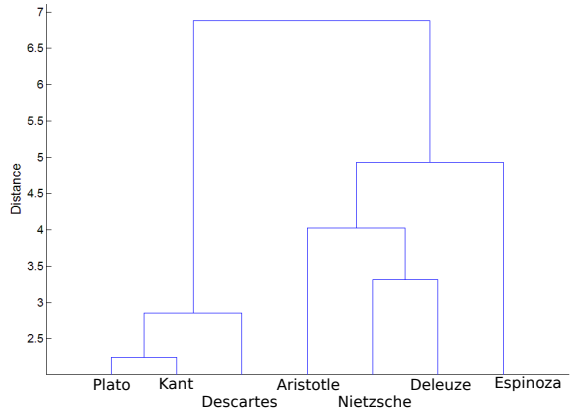}
        \end{center}
        \caption{\it Wards  hierarchical clustering of the seven
                    philosophers considering all the eight features.}
        \label{fig.hier}
\end{figure}

\section{Concluding Remarks}

Though science and philosophy share the same objective of explaining,
modeling, understanding and predicting nature, an essential difference
between them remains in that the latter has not relied systematically
on quantification of the natural world.  Several of the ambiguities of
the philosophical debate can be ultimately identified as consequences
of such a lack of quantitative rigour.

In this work, we reported a quantitative approach to characterize and
analyze philosophical thought based on the assignment of scores to
several philosophers regarding eight principal attributes that have
characterized philosophical thought.  In doing so, it became possible
to apply several concepts from multivariate statistics and pattern
recognition areas, and derive insights on the nature of philosophical
thinking along its historical evolution.  The high Pearson correlation
coefficients between the eight selected philosophical features
revealed strong interrelationships among them, which means that they
tend to go together, possibly even implying one another.  Further
studies would be however required to try to identify causality
relationships between these characteristics.

The application of Principal Components Analysis yielded additional
insights.  Confirming the high level of correlations between the
chosen philosophical characteristics, we found that the dispersion of
the philosophical states can be effectively projected into a
2-dimensional space, with several implications.  The apparent
complexity of philosophy as suggested by the eight original axes seems
to be largely illusory, being a consequence of the intense
correlations between the quantified philosophical characteristics.
Furthermore, the evolution of philosophy appears to be constrained as
a consequence of this mainly planar organization of the philosophical
space, in the sense that the philosophical moves are mostly restricted
to a plane region. Therefore, the hypothesis that philosophers seek
difference and innovation seems to be fundamentally constrained by the
relatively narrow space where philosophical moves have taken place.

Special attention was also given to formalizing the analysis of the
time evolution of philosophy, which was performed thanks to the
proposal of the opposition, skewness and counter-dialectics indices.
We believe these indices can be used to quantify innovation along the
development of philosophy.  To our great surprise, we found that
innovation in most philosophical moves have been mostly a consequence
of opposition to the current philosophical state, with rather small
skewness.  Also surprising was the identification of strongly
dialectics component in most pairs of subsequent philosophical moves.
This suggests that dialectics has played indeed a key role in the
development of philosophical thought.

It should be acknowledged that the scores and choice of main
philosophical characteristics adopted in the current work are largely
arbitrary and could be substantially improved.  However, the
perturbation analysis performed in this work suggests that the effect
of non-systematic errors in assigning the scores does not seem to be
critical and has little overall impact on the conclusions we have
derived.  In addition, whatever the effects of the scores and choice
of main features, it should be emphasized that the proposed
methodology can be readily applied to expanded and enhanced sets of
scores or philosophical characteristics.  As a matter of fact, this is
perhaps the main contribution of this work, i.e.\ the proposal of a
sound, formal quantitative methodology which can lead to
comprehensive, objective insights about how philosophy has evolved
since its earliest origins.  It is also worth observing that this
methodology is not restricted to individual philosophers, and it can
be adapted to the investigation of philosophical schools, individual
pieces (e.g.\ books), or even of the works of the same philosophers
along distinct periods of time.  It is also possible to apply this
methodology to other areas such as arts and science.

Going back to the beginning of this work, we conclude that
innovation/difference in philosophical development, at least as
represented by the chosen eight measurements and seven philosophers,
may have been driven mostly by opposition moves and dialectics in a
rather narrow effective philosophical space.  These results beg the
question of how the philosophical space could be expanded, so as to
favor innovations.  We are hopeful that quantitative approaches such
as the one reported here may represent some first steps in that
direction.

\begin{acknowledgments}
Luciano da F. Costa thanks CNPq (308231/03-1) and FAPESP (05/00587-5)
for sponsorship. Renato Fabbri is grateful to CAPES and 
the Postgrad Committee of the IFSC.  The authors thanks Debora
Correa for providing the Ward's hierarchical clustering of the
philosophers.
\end{acknowledgments}

\appendix

\section{A Brief Explanantion of Principal Component Analysis (PCA)}

In plain words, PCA is a dimensionality reduction procedure performed
through axes rotation.  It operates by concentrating
dispersion/variance along the first new axes, which are denominated
the principal components.

The technique consists in finding the eigenvectors and eigenvalues of
the covariance matrix of the respective random vectors (i.e.\ the
vectors associated with each philosophical state). The eigenvalues
correspond to the variances of the new variables.  When multiplied by
the original feature matrix, the eigenvectors yield the new random
variables which are fully uncorrelated.

For a more extensive explanation of PCA, please refer to~\cite{Costa}
and references therein.

\nocite{*}
\bibliography{phi4}

\end{document}